\documentclass[10pt]{article}

\usepackage{a4wide}
\usepackage{array}
\usepackage{amsmath}

\begin{document}

\title{Agile Effort Estimation: Comparing the Accuracy and Efficiency of Planning Poker, 
Bucket System, and Affinity Estimation methods}

\author{Marko Po\v{z}enel*, Luka F{\"u}rst, Damjan Vavpoti\v{c}, Toma\v{z} Hovelja \\
Faculty of Computer and Information Science \\
University of Ljubljana, Ve\v{c}na pot 113, \\
1000 Ljubljana, Slovenia\\
}

\date{}

\maketitle

\noindent *corresponding author

\vspace{1.5cm}

\noindent \textbf{Original contribution available in International Journal of Software Engineering and Knowledge Engineering (IJSEKE)}:

\vspace{0.5cm}

\noindent \textbf{Po\v{z}enel M, F{\"u}rst L, Vavpoti\v{c} D, Hovelja T, Agile Effort Estimation: Comparing the Accuracy and Efficiency of
Planning Poker, Bucket System, and Affinity Estimation methods, International Journal of Software Engineering and 
Knowledge Engineering (First Published December 21, 2023), \copyright 2023
(World Scientific Journals). DOI: 10.1142/S021819402350064X }

\vspace{1.0cm}

\begin{abstract}
Published studies on agile effort estimation predominantly focus on comparisons of the accuracy of different 
estimation methods, while efficiency comparisons, i.e. how much time the estimation methods consume was not 
in the forefront. However, for practical use in software development, the time required can be a very important 
cost factor for enterprises, especially when the accuracy of different agile effort estimations is similar. 
In this study, we thus try to advance the current standard accuracy comparison between methods by introducing 
efficiency i.e. time it takes to use a method as an additional dimension of comparison. We conduct this comparison 
between three agile effort estimation methods that were not yet compared in the literature, namely Planning 
Poker, Bucket System and Affinity Estimation. For the comparison, we used eight student teams with 29 students 
that had to use all the effort estimation methods during the course where they had to finish a programming 
project in 3 weeks. The results indicate that after the students get used to using the different methods the 
accuracy between them is not statistically significantly different, however, the efficiency is. On average 
Bucket System and Affinity Estimation methods take half as much time as Planning Poker.
\end{abstract}

\vspace{0.5cm}

\section{Introduction \label{intro}}

Software development project failures are common even though increasing importance is placed on factors that lead 
to project success. In the last decade, the failure rate of software projects remained high \cite{Lai-22,Nizam-22}. 
One of the main reasons for project failures are unplanned changes that occur during the project \cite{Serrador-15}. 
Traditional models for planning and execution, where we make a comprehensive plan at the beginning, which is 
followed by the implementation phase, may not be optimal for most software development projects since it limits 
the ability of the project team to address unplanned challenges that they are facing successfully. For this reason, 
many software companies have shifted to Agile Software Development (ASD) \cite{Usman-17,Cohen-04}. Agile development 
received widespread interest \cite{Dingsoyr-14} and caused a paradigm shift in software engineering \cite{Rajlich-06}, 
establishing agile as the new standard approach to software development project management.

Agile development includes several practices like extreme programming, Scrum, Test-driven development, and Crystal 
family \cite{Gandomani-19}, from which the most known and regularly used are Scrum, Extreme Programming (XP) and 
Kanban \cite{Diebold-15,Versionone-17-11th,Anwer-17,Balve-17,Tavares-19}. Scrum is an iterative, incremental and 
empirical method used to manage software development projects \cite{Lopez-18}. It provides numerous benefits to the 
industry, like higher project overview, transparency, improved communication between all parties, more efficient 
teamwork, improved requirements understanding, planning, focus on project outcome and better expectations management 
between stakeholders \cite{Rubin-12,Lopez-17}. It has been widely adopted in countless software development 
companies \cite{Lopez-18,Mahnic-11,Zahraoui-15,Srivastava-17}.

One of the most important phases in the software development process is effort estimation. It helps to ensure that 
the end product will be implemented and delivered according to the plan. It facilitates more detailed understanding 
of the project's functionalities \cite{Usman-17,Mahnic-12}. If it is not done properly, it will affect the project 
and, finally, the customer. According to Standish Group's Chaos Report \cite{chaos-14}, many software manufacturing 
enterprises failed because of inappropriate estimation of user stories \cite{Chongpakdee-17}. In agile software 
development, effort estimation plays an important role in the phase of release and iteration planning \cite{Usman-17}. 
Methods like Planning Poker (PP) or expert judgment are usually used. Scrum also does not prescribe an estimation 
technique. However, agile methods usually propose the use of Planning Poker \cite{Grenning-02} as an estimation 
technique for user story estimation \cite{Lopez-17,Raith-13}. Planning Poker is one of the frequently used effort 
estimation techniques \cite{Lopez-18,Gandomani-19,Sudarmaningtyas-20}, while Story points are one of the most 
commonly used size metrics \cite{Usman-14,Choetkiertikul-18}.

Group estimation techniques, like Planning Poker, are widely used since they engage the whole team and bring together 
multiple expert opinions. Planning Poker is a consensus-based estimation method where user stories in the product 
backlog are individually assessed, with incentivizing discussion as a key element \cite{Mahnic-12}. When assessing a 
large product backlog using Planning Poker, the estimation process can take lots of time \cite{Ionescu-17,Pozenel-19}. 
Also, assessing individual stories one by one might obscure the overall product backlog view \cite{Pozenel-19}. 
Novel group estimation techniques like the Team Estimation Game, Bucket System \cite{Mallidi-21}, and Affinity 
Estimation \cite{woolf-14} were proposed by practitioners as an alternative to Planning Poker. In the literature, 
there is a general lack of papers that empirically investigate the strengths and deficiencies of proposed assessment 
methods. Mahni\v{c} et al. \cite{Mahnic-12} investigated the accuracy of Planning Poker estimates compared to expert 
judgement. Po\v{z}enel et al. \cite{Pozenel-19} compared Planning Poker and Team Estimation Game in terms of 
estimation accuracy, while the Bucket System and Affinity Estimation comparison to Planning Poker in terms of 
accuracy and efficiency remains to our knowledge to be investigated in any empirical study at all. To start addressing 
this gap in the literature, we believe it is worth investigating how these three methods compare within a typical 
agile project setting with a small product backlog and a smaller team in terms of accuracy and efficiency.

The purpose of this paper is thus to compare, through a quantitative study, Planning Poker to two alternative estimation 
techniques, Affinity Estimation (AE) and Bucket System (BS). As stated in the literature \cite{Usman-17}, Planning Poker 
is one of the most popular agile effort estimation techniques in agile effort estimation studies. However, to date, 
quantitative studies have been lacking that would investigate Affinity Estimation and Bucket System techniques in 
project settings, testing the accuracy and time efficacy of these methods compared to Planning Poker. This paper 
reports on the results of such a study in a graduate course with computer science and informatics students. We 
investigated how these three methods compare in terms of assessment accuracy and assessment time efficiency.

The remainder of the paper is organized as follows: Section \ref{rw} presents related work in agile estimation. Section 
\ref{estimmethods} presents agile estimation techniques used in our study. Study design, results, and evaluation are 
presented in Section \ref{study-design} and Section \ref{calculation} and discussed in Section \ref{discussion}. 
Validity concerns are addressed and described in Section \ref{limitations}. Section \ref{conclusion} presents the 
conclusion and directions for future work.

\section{Related work \label{rw}}

Agile development is dynamic at its core. As a result, effort estimation in ASD has to address the dynamic way 
the software is developed. It is challenging to track, maintain and assess the effort needed to complete the 
product under development \cite{Bilgaiyan-19}. In the literature, considerable effort has been made to improve 
effort estimation for ASD. There are numerous studies on agile software development \cite{Usman-14,Usman-15,Fernandez-20}. 
Most papers study popular agile effort estimation techniques (such as Planning Poker) to determine the effort 
needed to complete the project \cite{Bilgaiyan-19}. Subjective assessment-based techniques (i.e. Planning Poker, 
analogy, expert judgment) align well with the agile way of thinking, where more emphasis is placed on people 
and their interaction \cite{Usman-15}. These techniques require a subjective assessment by the experts \cite{Usman-15}. 
Estimates are based on team members' experiences and perceptions and can, therefore, be viewed as subjective. 
The assessment techniques used in ASD are usually simple and suitable for group assessment. The process and 
steps of Planning Poker or Team Estimation Game are simple and easy to apply \cite{Gandomani-19,Mahnic-12,Pozenel-19}.

On the other hand, questions have been raised in the literature about whether computational estimation methods 
could be used to solve the problem of effort estimation in ASD projects. Bilgaiyan et al. \cite{Bilgaiyan-19} presented 
an artificial neural network approach for effort estimation in ASD. Saini et al. \cite{Saini-18} employed a fuzzy 
logic model to estimate the effort of agile software projects. Effort estimation of agile projects using Machine 
Learning Techniques is presented in \cite{Prasada-18}. Dantas et al. \cite{Dantas-18} presented a systematic 
literature review on effort estimation in ASD, including computing techniques. However, they found no evidence 
of the benefits of computing techniques and whether the methods were validated in the industry. We believe that 
additional subjective assessment methods are worth exploring. This paper focuses on subjective assessment-based 
techniques.

We compare two cost assessment methods, Affinity Estimation \cite{Mallidi-21,woolf-14} and Bucket System \cite{Mallidi-21}, 
with Planning Poker. In the literature, neither Affinity Estimation nor Bucket System (also known as a T-shirt 
estimation) is a novel assessment method and has been addressed in the papers before \cite{Mallidi-21,Abadeer-21}. 
In addition, practitioners have been using them for years \cite{woolf-14,affinity-18}. However, based on the 
conducted literature review, they have not been empirically compared in an empirical study before. We believe 
that this is the first work comparing Planning Poker, Affinity Estimation and Bucket System in terms of required 
time and assessment accuracy.

Planning Poker is a group estimation method based on team consensus \cite{Gandomani-14}. It is briefly presented 
in Section \ref{estimmethods} and is described in more detail in Mahni\v{c} et al. \cite{Mahnic-12}. It is one of the 
most popular software estimation techniques in ASD \cite{Gandomani-19,Canedo-18}. What is more, Altaleb et al. 
\cite{Altaleb-20} found that Expert Judgment and Planning Poker remain one of the most popular estimation 
techniques in mobile app development as well. In their paper, Alhamed and Storer \cite{Alhamed-21} state that 
Planning Poker can produce reliable estimates with knowledgeable domain experts. The challenge of Planning Poker 
is that it can be time-consuming in practice and can have difficulties with reaching an agreement in time 
\cite{Gandomani-19,Tanveer-17}.

In their paper, Mallidi and Sharma \cite{Mallidi-21} presented various assessment techniques and their challenges 
in agile development. In the estimation phase, Planning Poker focuses on individual user stories, which can 
result in a high cognitive load when assessing a large number of user stories. When there is a large number of 
user stories to be estimated, the collaborative method, the Bucket System, is more appropriate \cite{Mallidi-21}. 
It enables a non-time-consuming estimation of many user stories with a medium to big-sized group of people 
\cite{Yuliansyah-18,Waldeck-20,agileadvice-bs}. It is time efficient but needs an experienced team for proper 
assessment \cite{Mallidi-21}.

Affinity Estimation (also known as Affinity Mapping) is another popular collaborative method for effort estimation 
used in agile projects. In the literature, Affinity Estimation is recognized as a technique used for effort 
estimation in agile development \cite{Mallidi-21,Venkata-17,Sedano-20}. At the time of writing, Affinity Estimation 
and Bucket System are among the top 7 agile estimation techniques according to blog z-stream \cite{zstream-ae}. 
Sedano et al. \cite{Sedano-20} define Affinity Estimation as "a lightweight, qualitative data analysis technique 
in which the team spreads notes on a large table or wall and collectively arranges related notes into groups". 
As with the Bucket System, Affinity Estimation is a relative estimation technique focusing on grouping stories 
that match \cite{zstream-ae}. It enables the team to estimate a large set of stories in a concise amount of 
time. A shortcoming may present quite a brief discussion around the user stories, so we can miss out on valuable 
information that could affect the user story assessment score \cite{Venkata-17}. Practitioners find Affinity 
Estimation simple, transparent, and less time-consuming, reducing cognitive load \cite{parabol-ae} and making 
the estimating session a joyful and collaborative experience \cite{amond-22}.

Another relative estimation method used in agile software development is the Team Estimation Game, described in 
more detail in Po\v{z}enel and Hovelja \cite{Pozenel-19}.

The use of capstone courses \cite{Umphress-02} in software engineering education is not new, but in the last 
two decades, they have been gaining more relevance \cite{Bastarrica-17}. However, because of external validity 
concerns, practitioners and the science community often view Empirical Studies with Students (ESWS) skeptically 
\cite{Carver-04,Carver-10,Falessi-18}. To address the concern, Carver et al. \cite{Carver-10} presented constraints, 
goals and guidelines to fully exploit empirical studies with students to fight validity threats caused by the 
involvement of students as the participants. Carver et al. \cite{Carver-04} also introduced a checklist to 
ensure that empirical studies with students would provide as much research and pedagogical value as possible. 
However, in their study, Falessi et al. \cite{Falessi-18} report that software engineering experiments were 
many times judged as having severe threats to validity because they used students as subjects. Nevertheless, we 
believe empirical studies with students can provide value to the industrial and research community if the study 
design and implementation comply with the requirements addressed in the paper presented by Carver et al. 
\cite{Carver-04}. In addition, ESWSs can also be beneficial for students and instructors \cite{Carver-04}.

A systematic literature review on effort estimation in ASD is presented in the paper, analysed by Fernández-Diego 
et al. \cite{Fernandez-20}. Among these studies, we highlight a capstone project that investigates the performance 
of expert-based assessment methods in agile development. In  Mahni\v{c} et al. \cite{Mahnic-12}, the authors 
compared the accuracy of Planning Poker with a statistical combination of individual experts' estimates and 
found that Planning Poker provides more accurate estimates. Po\v{z}enel and Hovelja \cite{Pozenel-19} conducted 
a study that assesses the accuracy of two popular assessment methods, Planning Poker and Team Estimation Game. 
They showed that the Team Estimation Game provides more accurate story estimates than Planning Poker. 
Nevertheless, they were unable to show that the Team Estimation Game is less time-consuming than Planning Poker.

\section{Group estimation methods\label{estimmethods}}

Planning Poker is used to estimate user stories at the estimation session, where the development team estimates 
user stories' complexity one by one. Each team member receives one deck of evaluation cards presenting a 
predetermined ascending sequence of numbers, usually the Fibonacci sequence.

The assessment process begins when one of the Team members (or the Scrum Master) takes the first user story card 
from the Product Backlog. The Product Owner presents the user story requirements. The development team members may 
ask questions to the Product Owner and discuss the user story details among themselves. To avoid anchoring, team 
members should not discuss story estimates directly. All team members reveal their estimates at the same time after 
the discussion. If their assessment differs too much, the team discusses their estimates. The team members with high 
and low estimates explain reasons for their story estimate. After the discussion, they repeat the estimation process 
by revealing the evaluation cards. Usually, the estimates start to converge after the second round. We repeat the 
process until a consensus is reached. The assessment is complete when the team has assessed all the user story 
cards in the Product Backlog. Planning Poker and its steps were presented by Grenning \cite{Grenning-02}.

The following describes the Affinity Estimation and Bucket System group estimation methods.

\subsection{Affinity Estimation}

The Product Owner provides a product backlog while the development team prepares a working space for placing the 
story cards. At the top of the assessment area, prepare a horizontal scale from "smaller" to "larger". The Product 
Owner distributes all story cards from the product backlog to the development team members. In the relative sizing 
phase, team members estimate the effort of assigned user stories without discussion with other team members. User 
stories have to be placed close to each other, considering the effort needed for completion. The Product Owner is 
present during an assessment to clarify user stories. User stories that team members  are unable to correctly place, 
even after discussing the Product Owner, are put in a separate place outside the assessment area. At the end of the 
relative sizing phase, user stories are arranged on the assessment board from smallest to largest. User stories 
stacked vertically are about the same relative size.

The silent phase is followed by the assessment board alteration phase, where team members edit the relative sizes 
of story cards if necessary. They have to involve other team members in the discussion before making major moves. 
The team rearranges user stories according to new findings. Toss-ups are discussed and clarified with the Product 
Owner. This phase continues until there are only a few changes on the board.

Once user stories have been edited, the team members place them into size buckets. User stories are already arranged 
on a scale from smaller to larger. The team creates buckets and places their sizes along the spectrum at the top of 
the assessment area. Different nomenclatures can be used, from a Fibonacci sequence ($1/2$, $1$, $2$, $3$, $5$) to 
T-Shirt sizing. The bucket size marks are placed in positions relative to their size. The larger the bucket, the 
greater the distance between the marks. Team members draw lines between buckets to make the arrangement between 
buckets visible. There can be several user story columns in each bucket, while user story cards may be unevenly 
distributed between buckets.

The next step is the Product Owner's "challenge". In this step, the Product Owner selects a set of user stories 
from the board and discusses their size estimates with the development team. The development team discusses the 
reasons for assigning size estimates associated with each selected user story. If the discussion results in changes 
in story size estimates, the stories in question move to new buckets. This step is complete when all selected user 
stories have been discussed.

All that remains is to write story estimates in the product backlog of the agile project management tool 
\cite{affinity-18,techagilist-ae}.

\subsection{Bucket System}

Before starting, the Product Owner develops a product backlog while the development team prepares a working 
space to place user story cards during the assessment. At the top of the assessment space, the team places cards 
with effort markers from the smallest to the largest. Cards with effort markers are called buckets. Different 
nomenclatures can be used. However, the Fibonacci series is recommended.

One of the team members picks a random user story from the product backlog. The team discusses it and places it 
in the centrally placed bucket. For instance, if the Fibonacci series ends with the number $34$, put it in the 
bucket marked with the number $4$. If the selected story happens to belong to one of the edge buckets, the team 
replaces it with another whose effort places it in a more central bucket. The first user story serves as a 
reference for the other user stories. In the process, the team estimates user stories' effort, each relative to 
the others.

The team randomly picks two more user stories from the product backlog, discusses the details of the selected 
two stories and compares them against the reference user story. All aspects of development, from the user interface, 
business logic, testing and integration, must be considered. Once consensus has been met, the team puts selected 
two story cards in the appropriate bucket. After the first three user stories are assessed and bucketed, the 
team checks the position for skewness towards one of the two ends of the scale. If necessary, the team removes 
skewness by repositioning user stories or adjusting the sizes and range of the buckets.

The divide and conquer phase follows the sampling phase. The Product Owner distributed the remaining user stories 
from the product backlog among the development team members. Each team member places his or her user stories in 
the buckets independently, without any discussion. If a team member does not understand a user story, he can 
exchange it with the story of another team member. The exchange should not lead to a discussion.

When all user stories are distributed to the buckets, the development team does a quick, silent review of the 
stories' layout. If a team member disagrees with the position of a particular card, they notify the team, which 
then initiates a discussion until reaching a consensus. Scrum Master assures that only now can cards be moved. 
This phase, called "sanity check", is a crucial phase in the assessment process.

The last step is writing bucket numbers on the story estimates, then copying story estimates to the product 
backlog of the agile project management tool. The assessment is complete \cite{agileadvice-bs,techagilist-bs}.

\section{Study Design\label{study-design}}

Given the problematics presented in the abstract and introduction, we believe that an empirical study including 
both accuracy and efficiency assessments of agile estimation methods will benefit agile software development in 
business and pedagogical settings, wherever time constraints are of significance. To the best of our knowledge, 
a comparison of the accuracy and efficiency of different agile estimation methods has not been published yet. 
Similarly, a comparison between Planning Poker, Bucket System and Affinity Estimation methods is also still 
missing from the literature. Thus, academics and practitioners should be interested in the results of this study. 
To be able to statistically compare estimates of different project teams and produce relevant results, all eight 
student teams had to complete the same software project with the same set of 23 “must-have” project user stories. 
Since it is almost impossible to find a professional industrial setting where multiple teams would be working on 
the same set of project user stories, using students as test subjects was a way to bypass its limitations of 
professional settings and proceed with the study. We used a within-subjects design with 29 participating students.

\subsection{Research questions}

Our study aimed to evaluate the accuracy and efficiency of the three above-presented agile estimation methods. 
In order to achieve this aim, we posed the following research questions that are based on relevant literature:

\begin{itemize}

    \item RQ1: Is any of the three studied agile estimation methods more or less accurate than the other two \cite{Mahnic-12,Molokken-08}?
		The null hypothesis for RQ1 is: there is no statistically significant difference in accuracy of estimated 
		effort to complete project user stories between studied agile estimation methods of Planning Poker, Bucket 
		System and Affinity Estimation. The alternative hypothesis is: certain agile estimation methods are statistically significantly 
		more accurate than others in accuracy of estimated effort to complete project user stories.
		
    \item RQ2: Is any of the three studied agile methods more or less time efficient than the other two \cite{Pozenel-19}? 
		The null hypothesis for RQ2 is: there is no statistically significant difference in time needed to estimate 
		project user stories in each Sprint between studied agile estimation methods of Planning Poker, Bucket System 
		and Affinity Estimation. The alternative hypothesis is: certain agile estimation methods are statistically significantly more 
		efficient than others in time needed to estimate project user stories in each Sprint.
		
\end{itemize}
RQ1 will explore how accurate the three agile estimation methods are compared to each other. This research 
question is usually central in comparable studies of different agile estimation methods \cite{Mahnic-12,Venkata-17,Molokken-08,Molokken-03,Mahnic-11capstone}. 
RQ2 will explore the time efficiency of the studied methods. As discussed in the relevant work section, Planning 
Poker is very popular \cite{Gandomani-19,Lopez-18,Chongpakdee-17,Canedo-18}, even though several authors point 
out how time-consuming it is \cite{Gandomani-19,Tanveer-17,Bjorsne-13}. If Bucket System and Affinity Estimation 
turn out to consume significantly less time than Planning Poker with similar estimation accuracy, then academics 
(what we teach) and practitioners alike (what we use in practice) should start to strongly consider these two 
currently less used alternative effort estimation methods.

\subsection{The major Project course}

Modern Methods of Software Development course is designed to teach agile \cite{Cohen-04} and lean \cite{Poppendieck-03} 
software development methods. The course lasts 15 weeks and is included in the first year of the master's curriculum 
in the Computer and Information Science program at the University of Ljubljana and brings six European Credit Transfer 
and Accumulation System (ECTS) points. Therefore, it requires a student's workload between 150 and 180 hours 
(including contact hours). In the last year of undergraduate studies in the Computer and Information Science program, 
students focus on mastering traditional methods of software development. At the same time, less attention is paid to 
agile development. Thus, this postgraduate course is dedicated to mastering agile software development methods like 
Scrum. In addition, students acquire knowledge about lean manufacturing concepts. In particular, the course focuses 
on the Scrumban \cite{Ladas-09} and Kanban methods as typical representatives of lean trends.

The survey on the State of agile in Slovenia in 2019 shows that the most widespread and used agile methods are Scrum (62\%) 
and Kanban (47\%) \cite{Svetelj-19}. The survey states that Slovenian software companies are increasingly focusing on agile. 
Similarly, the State of Agile global survey in 2022 shows a similar lead in popularity for Scrum (87\%) and Kanban (56\%) over 
other agile methodologies, followed by Scrumban (27\%) \cite{digitalai-22}. Scrumban has grown from 10\% in the 2020 survey 
to 27\% in the 2022 survey. Kanban and Scrum practitioners are thus increasingly combining both methods \cite{Alqudah-18}. 
Alqudah et al. \cite{Alqudah-17} state that minimizing waste and the likelihood of project delay, improving the quality of 
the delivered products, and the ability to make improvements continuously are the main benefits of Scrumban \cite{Ladas-09}. 
Scrumban, as a hybrid of the two most common methods in workplaces in Slovenia, thus enables students to familiarize themselves 
with concepts of both most dominant methods in their future workplace. For this reason, we found it sensible to feature 
it prominently in the curriculum.

The application of the methods is demonstrated through practical work, which serves as a case study for the empirical 
evaluation of the development process and its results. Students are required to work in groups to develop a quasi-real 
project based on the Product Owner's requirements. Through the capstone project \cite{Umphress-02}, students gain 
experience estimating agile project complexity and tracking and managing the agile project. In addition, they acquire 
additional skills, such as the ability to work in an agile team, organize and lead, communicate with the Product Owner 
and present the results to the Product Owner.

\subsection{An overview of the ESWS}

Carver et al. \cite{Carver-10} have built on their experience conducting ESWS and identified requirements for a successful 
ESWS. Based on these requirements, they proposed a checklist to help researchers to make ESWS as effective as possible. 
Checklist items help researchers to address pedagogical and study goals. To study the differences between group estimation 
methods, we designed and incorporated ESWS into the capstone course according to the checklist. Table 1
provides a high-level overview of the checklist items.

\renewcommand{\arraystretch}{1.2}

\begin{table}[h!]
\small
\begin{center}
\begin{tabular}{ r l }

\hline\hline 
\multicolumn{2}{l}{Before the class begins} \\ 
\hline 
1  &  Ensure adequate integration of the study into the course topics. \\
2  &  Integrate the study timeline with the course schedule. \\
3  &  Reuse artefacts and tools as appropriate. \\
4  &  Write up a protocol and have it reviewed. \\
\hline
\multicolumn{2}{l}{As soon as the class begins} \\
\hline
5  &  Obtain subjects' permission for their participation in the study. \\
6  &  Set subject expectations. \\
\hline
\multicolumn{2}{l}{When the study begins} \\
\hline
7  &  Document information about the experimental context in detail. \\
8  &  Implement policies for controlling/monitoring the experimental variables. \\
\hline
\multicolumn{2}{l}{When the study is completed} \\
\hline
9  &  Plan follow-up activities. \\
10 &  Build or update a laboratory package. \\
\hline\hline

\end{tabular}
\caption{ESWS checklist~\protect\cite{Carver-10}.}
\label{table_esws_checklist}
\end{center}
\end{table}

To ensure adequate integration of the study into the course topics, we ensured that the students who participated 
in the ESWS had the necessary knowledge and that the research goals fit into the class materials. The study was 
conducted with students of the software engineering module and took place in the first semester of their post-graduate 
study. Within Sprint 0, students were thoroughly briefed on the tools, Kanban Scrum and Scrumban methods, and 
assessment methods under study. Special care was taken to ensure that pursuing the research goal did not hinder 
teaching goals. The user story assessment process is an integral part of the capstone course. Study execution did 
not require additional student effort except faithful recording of collected data.

Scrum and Scrumban require effort estimation based on user stories at the beginning of each Sprint. Each user story 
must be further decomposed into tasks, and the time required to complete each task must be estimated. Integration of 
the study into the course schedule did not require any adaptation of the course process. At the beginning of the 
course, we defined the time scale at which the participants were required to collect data. Students recorded story 
and task estimates during the Sprint planning meeting for each Sprint, and the time spent to complete the tasks was 
recorded when the task or user story was complete. The team and teaching staff verified faithful data logging at 
each daily Stand-up meeting.

Regarding reusing Existing Artefacts, we drew on experience with past ESWS. We conducted a similar ESWS in the 
capstone course in the final year of undergraduate study \cite{Mahnic-12,Pozenel-16,Pozenel-19}. A standard tool, 
Kanbanize, has been used for managing students' projects and facilitating empirical data collection. Since the 
automatic collection of story estimates during the assessment process is not supported in the Kanban, we prescribed 
precise procedures for when and how the participants recorded the data, the organisation of the Scrumban project 
board, and custom attributes of the cards presenting user stories. In each student team, one student played a 
Kanban Master role responsible for appropriately using the Kanbanize tool and process consistency with the 
methodology.

Concerning the protocol revision, we developed the protocol based on the Scrumban framework and course requirements. 
Faculty colleagues with previous experience with ESWSs reviewed the protocol, and their recommendations were followed. 
According to the protocol, we defined activities in initial Sprint 0, the following four sprints and at the course end. 
In Sprint 0, students learned Kanban, Scrum, Scrumban, and all assessment methods during classical lectures. In order 
to simulate a real environment, the course requires working in teams, so students had to form teams of four members and 
prepare a development environment. During the lab sessions, they were trained in managing their project using the 
Kanbanize tool and practicing Planning Poker, Affinity Estimation and Bucket System estimation methods. After the 
students were acquainted with the Product Backlog and fully understood the course project requirements, they were 
asked for consent for their participation in the study. The teams that participated in the study had no additional 
work compared to others. The capstone course had four regular Sprints (1, 2, 3, 4), where the teams were required 
to re-estimate unfinished stories, decompose them into tasks, and record the amount of work done. Daily stand-up 
meetings were used to monitor data collection and compliance with protocols. At the end of the course, the researchers 
and instructors performed a statistical analysis of the collected data. The instructors presented preliminary study 
results to the students during the last lecture.

As for obtaining subjects' permissions, we asked the students for their permission to participate in the research 
after they acquired information about the study course, course obligations, and Product Backlog and obtained sufficient 
information about the required methodology and assessment methods. We explained that participation is voluntary and 
stressed that there is no difference in effort needed between the teams that participate and those that do not. We 
emphasized the importance of conscientiously performing activities and recording data.

Concerning subjects' expectations, we tried to ensure that the students focused on the learning goals. Although we 
wanted to motivate students to participate in the study, we emphasized that they should not feel obligated to 
participate. To increase motivation to perform tasks as conscientiously as possible, each participating student 
got up to ten additional points on the exam, depending on conscious task execution and quality data logging. In order 
not to hinder the study results, the students were only roughly informed about the study goals. Their expectations 
were mainly focused on teaching goals, where they learned agile software development practices through the practical 
project. We did not discuss the study details with the students, who might deliberately manipulate the results to 
confirm the research hypotheses.

We created the document with the experimental context as a basis for possible study replication. The documents 
describe teaching goals, topics covered, teaching methods, and research questions.

Concerning policies for monitoring the experimental variables, we incorporated the study within the project work 
seamlessly without hindering students' work on their projects. Effort estimation took place on-site under the supervision 
of instructors and the Product Owner, offering details about project requirements. At Daily stand-up meetings, 
instructors monitored data collection, allowing them to draw attention to possible deficiencies. The project management 
tool Kanbanize allowed students to collect data on their projects seamlessly.

Follow-up activities included calculating the accuracy of estimates using a balanced measure of relative error. We 
analyzed data about estimates and time of estimation between Planning Poker, Affinity Estimation and Bucket System 
using well-known statistical methods, such as the Mann-Whitney-Wilcoxon test. We did not provide a formal lab package 
allowing study replication out of the box. However, we have documented all the steps for creating such a package.

\subsection{Project setting}

Various estimation methods are used in Agile Software Development, from expert judgment to Artificial Intelligence-based 
methods \cite{Usman-14} and even traditional methods \cite{Osman-16}. However, the most used are those that use 
expert-based subjective assessment \cite{Usman-14}. Planning Poker has been used in the software industry and teaching 
environment \cite{Labedzki-17,Gandomani-19,Chatzipetrou-18,Lopez-17}. It can provide reliable estimates compared to 
single expert estimation \cite{Alhamed-21}. Labedzki et al. \cite{Labedzki-17} presented a study based on three 
different reference agile projects with a mature team, different team sizes, project lengths and user stories that 
differed in size and how they were defined. They concluded that Planning Poker yielded good results in any of the 
three settings, improved estimation precision and enabled a better understanding of user stories.

Unlike Planning Poker, Bucket System and Affinity Estimation are based on relative estimating. Mallidi et al. \cite{Mallidi-21} 
proposed Bucket System for a larger number of backlog items with an experienced team for estimation. In comparison, 
how successful the method would be in projects with a relatively smaller number of developers, fewer backlog items, 
and a less experienced team remains till this study empirically untested. Additionally, to our knowledge, Planning 
Poker, Bucket System, and Affinity Estimation have yet to be compared in any empirical study. To start addressing 
this gap in the literature, we believe it is worth investigating how these three methods compare within agile project 
settings with smaller teams and product backlogs in terms of accuracy and efficiency.

The project the students were tasked with was the development of a web-based Kanban project management tool. During 
the lectures, students thought about the Kanban method and were presented with a Kanban project management tool, 
Kanbanize \cite{Mahnic-18}. Students then used this tool to manage their Kanban project, while the teaching staff 
defined realistic user requirements and took over the Product Owner and Kanban Master roles. All student teams were 
working on the same project with the same set of project user stories. Our study setup involved eight student teams, 
each consisting of four members and acting as a self-organizing and self-managing Kanban team. In the initial Sprint 
(Sprint 0), students prepared the development environment and got acquainted with the project management tool 
Kanbanize, project and study requirements and product backlog. The rest of the course was divided into four Sprints, 
each lasting three weeks.

At the beginning of each Sprint, the Product Owner selected user story cards for the sprint backlog in the next 
Sprint. All development teams started a new Sprint with the same set of user story cards in the Sprint backlog. 
Based on the experience with capstone projects in the past, the Product Owner set such a card workload that all 
student teams were able to complete all the story cards from the sprint backlog. At the beginning of each Sprint, 
teams were asked to estimate the effort to complete user stories assigned to that Sprint with all three estimation 
methods under study. The order in which the methods were used constantly changed and differed as much as possible 
between Sprints and between teams, so each team did as many permutations of the order in which the methods were 
used as possible (Table 2). During the Sprint, teams had regular Daily Stand-up meetings twice a week and maintained 
their Sprint backlogs in the Kanbanize tool. At the end of each Sprint, students present their results to the Product 
Owner at the Sprint review meeting and discuss improvements for the next Sprint at the retrospective meeting.

Jedlitschka et al. \cite{Jedlitschka-05} suggest that each controlled experiment in Software Engineering should 
clearly define and report the variables. Independent variables in our empirical study are 
i) the type of method to estimate effort to complete project user stories: Planning Poker, Affinity Estimation and 
Bucket System, and 
ii) the Actual effort to complete each project user story. Dependent variables are i) estimated effort to complete 
a user story for each estimation method and ii) time needed in each Sprint to estimate effort to complete user 
stories by each method. 
No controlled or random variables have been involved. In order to mitigate the confounding factor of maturation, 
we made sure that each time students used assessment methods, they were used in a different sequence. 
\begin{table}[h!]
\small
\begin{center}
\begin{tabular}{ p{1cm} c c c c c c c c }

\hline\hline 
          &  Team\,1  &  Team\,2  &  Team\,3  &  Team\,4  &  Team\,5  &  Team\,6  &  Team\,7  &  Team\,8 \\					
\hline
 & \multicolumn{8}{c}{Sprint 1} \\
\hline
          &  PP  &  PP  &  BS  &  BS  &  AE  &  AE  &  PP  &  PP \\
Sequence  &  BS  &  AE  &  PP  &  AE  &  PP  &  BS  &  BS  &  AE \\
          &  AE  &  BS  &  AE  &  PP  &  BS  &  PP  &  AE  &  BS \\
\hline
 & \multicolumn{8}{c}{Sprint 2} \\
\hline
         &  AE  &  BS  &  AE  &  PP  &  BS  &  PP  &  AE  &  BS \\
Sequence &  PP  &  PP  &  BS  &  BS  &  AE  &  AE  &  PP  &  PP \\
         &  BS  &  AE  &  PP  &  AE  &  PP  &  BS  &  BS  &  AE \\
\hline
 & \multicolumn{8}{c}{Sprint 3} \\
\hline
         &  BS  &  AE  &  PP  &  AE  &  PP  &  BS  &  BS  &  AE \\
Sequence &  AE  &  BS  &  AE  &  PP  &  BS  &  PP  &  AE  &  BS \\
         &  PP  &  PP  &  BS  &  BS  &  AE  &  AE  &  PP  &  PP \\
\hline
 & \multicolumn{8}{c}{Sprint 4} \\
\hline
         &  PP  &  PP  &  BS  &  BS  &  AE  &  AE  &  PP  &  PP \\
Sequence &  AE  &  BS  &  AE  &  PP  &  BS  &  PP  &  AE  &  BS \\
         &  BS  &  AE  &  PP  &  AE  &  PP  &  BS  &  BS  &  AE \\

\hline\hline
\end{tabular}
\caption{A sequence of estimation methods used in each Sprint.}
\label{table_method_sequence}
\end{center}
\end{table}

To maximize the research and teaching goals, we carefully designed and integrated the ESWS into the presented 
capstone course based on Carver's checklist \cite{Carver-10} and our past hands-on experience of conducting ESWSs 
\cite{Mahnic-12,Pozenel-19}.

\section{Calculation of Accuracy and Efficiency Estimation\label{calculation}}

To be able to answer the two stated research questions, we have to first define how accuracy and time efficiency 
will be measured. To measure accuracy, we will use BRE as it is the standard accuracy measure currently used in 
the literature \cite{Mahnic-12,Pozenel-19,Molokken-08,Molokken-07,Molokken-07rel,Molokken-05}. BRE (Balanced Relative 
Error), as shown in Eq. (1), is a measure of accuracy estimation that Miyazaki et al. \cite{Miyazaki-91} proposed 
in order to address the issues of the accuracy measure MRE (Magnitude of Relative Error) that was used in the 
literature prior to BRE gaining popularity.
\begin{equation}
BRE = \frac{\mid \text{actual effort} - \text{estimated effort} \mid}
{
	\min{(\text{actual effort}, \text{estimated effort})} 
}\, .
\label{eq:bre}
\end{equation}

BRE solves the main concern when using MRE, which is the uneven (non-balanced) weighting of underestimates and 
overestimates \cite{Foss-03}. The estimated effort was reported by student teams at the start of each Sprint for 
all the unfinished project user stories in the backlog. The students had to thus use all three estimation methods 
under examination each Sprint (every time in a different order). The actual effort is self-reported by student teams 
at the end of the Sprint in which they completed a certain user story. Similarly, student teams also self-reported 
how much time it took (minutes used per method) to estimate the time each of the studied methods consumed in each 
Sprint. Since students completed six user stories in every Sprint, we have 48 BRE estimates of accuracy in each 
Sprint but only eight-time efficiency measures for each studied method for each Sprint. Studied estimation methods, 
namely, were used only once by each team in each Sprint, as is standard in the agile methodology. Thus, due to 
sample size requirements, time estimation analysis couldn't be done on the level of individual Sprints as was the 
case for accuracy analysis but only for all four Sprints combined (32 time used self-reports for each method). 
Because BRE values had skewness larger than one and/or kurtosis larger than two, we rejected the assumption of 
normal distribution of our data \cite{Pozenel-19} and used non-parametric tests to answer our research questions. 
Because the two posed research questions required the comparison of two independent samples (BRE, time used of one 
studied estimation method vs BRE, time used of another studied estimation method), we used the non-parametric 
Mann-Whitney-Wilcoxon test \cite{Mann-47} as our main tool of analysis. With this tool, we searched for statistically 
significant differences between the three studied methods (P-value lower than 0.05) in BRE's and used time.

To ensure that statistically significant differences also have adequate effect size values, we used a standardized 
non-parametric effect size measure, Cliff's Delta. Cliff's Delta is a more powerful and robust measure than 
Cohen's d under certain conditions like skewed marginal distributions \cite{Kromrey-98analysis,Macbeth-11cliff}, 
and it works well for small to moderate samples ($n > 10$) \cite{Goedhart-16calculation}. Based on the categories 
first defined by Cohen, Vargha and Delaney \cite{Vargha-00critique}, calculated that Cliff's Delta effect sizes 
of $0.11$, $0.28$ and $0.43$ correspond to small, medium and large effects, respectively \cite{Goedhart-16calculation}.

\begin{table}[h!]
\small
\begin{center}
\begin{tabular}{ lllcclclcl }
\hline\hline 

	 & Method     & N	& Mean  &	Median     & Std.      & Skewness  & Std.\,Err.  & Kurtosis	& Std.\,Err. \\ 
	 & accuracy   &   &       & 	         & Dev. &           & of Skewness &          & of \\
	 &            &   &       &            & $\sigma$  &           &             &          & Kurtosis \\
\hline
					 & BRE pp & 48 & 1.23 & 0.66 & 1.82 & 3.67 & 0.34 & 17.55\hphantom{0} & 0.67 \\ 
	Sprint 1 & BRE bs & 48 & 1.74 & 0.90 & 2.26 & 2.31 & 0.34 & 6.05 & 0.67 \\ 
					 & BRE ae & 48 & 1.72 & 1.01 & 1.96 & 1.76 & 0.34 & 2.37 & 0.67 \\ 
\hline
					 & BRE pp & 48 & 1.24 & 0.71 & 1.50 & 2.08 & 0.34 & 4.44 & 0.67 \\
	Sprint 2 & BRE bs & 48 & 1.40 & 0.61 & 2.12 & 2.95 & 0.34 & 9.96 & 0.67 \\
					 & BRE ae & 48 & 1.67 & 0.87 & 1.96 & 2.73 & 0.34 & 10.17\hphantom{0} & 0.67 \\
\hline				
					 & BRE pp & 48 & 0.77 & 0.55 & 0.75 & 1.88 & 0.34 & 4.13 & 0.67 \\
	Sprint 3 & BRE bs & 48 & 0.99 & 0.63 & 1.02 & 1.62 & 0.34 & 2.00 & 0.67 \\
					 & BRE ae & 48 & 1.03 & 0.63 & 1.13 & 2.25 & 0.34 & 6.26 & 0.67 \\
\hline	
					 & BRE pp & 48 & 1.28 & 0.85 & 1.38 & 2.28 & 0.34 & 6.03 & 0.67 \\
	Sprint 4 & BRE bs & 48 & 1.58 & 0.88 & 1.95 & 2.90 & 0.34 & 11.07\hphantom{0} & 0.67 \\
					 & BRE ae & 48 & 1.55 & 1.10 & 1.54 & 2.13 & 0.34 & 5.27 & 0.67 \\

\hline\hline
\end{tabular}
\caption{Basic descriptive statistics for statistical testing of accuracy of estimated effort to complete project 
user stories between studied agile estimation methods of PP, BS and AE.}
\label{table_basic_descriptive_stats_1}
\end{center}
\end{table}

Table 4 presents the data on time used in minutes by student groups for a single use of the three studied 
methods for all four Sprints.

\begin{table}[h!]
\small
\begin{center}
\begin{tabular}{ lccclclcl }
\hline\hline 
						& N & Mean & Median & Std.      & Skewness & Std.\,Error  & Kurtosis & Std.\,Error \\
						&   &      &        & Dev.      &          & of Skew- &          & of         \\
						&   &      &        & $\sigma$  &          &  ness           &          & Kurtosis   \\
	\hline	
	Time used PP & 32 & 27.2188 & 24.5000 & 14.09583 & 1.548 & 0.414 & \hphantom{-}2.806 & 0.809 \\
		 (minutes) &    &         &   &   &   &   &   &  \\
	Time used BS & 32 & 16.5313 & 15.0000 & \hphantom{0}6.72973 & 0.488 & 0.414 & -0.371 & 0.809 \\
		 (minutes) &    &         &   &   &   &   &   &  \\
	Time used AE & 32 & 16.0000 & 16.0000 & \hphantom{0}8.03219 & 1.129 & 0.414 & \hphantom{-}1.836 & 0.809 \\
		 (minutes) &    &         &   &   &   &   &   &  \\
\hline\hline
\end{tabular}
\caption{Basic descriptive statistics for statistical testing of time needed to estimate project user stories 
in each Sprint between studied agile estimation methods of PP, BS and AE.}
\label{table_basic_descriptive_stats_2}
\end{center}
\end{table}

\subsection{Testing the first H0: There is no statistically significant difference in accuracy of estimated 
effort to complete project user stories between studied agile estimation methods of PP, BS and AE.}

We compared all three studied estimation methods to each other in each Sprint separately and also for all 
Sprints in aggregate. As shown in Table 5, the only statistically significant difference in accuracy is for 
the combined data for all Sprints between the Planning Poker and Affinity Estimation, where Affinity Estimation 
is statistically significantly less accurate than Planning Poker. The effect size measure of this statistical 
difference is small ($0.15$). For Sprints 1, 3 and 4, Planning Poker has a non-statistically significantly better 
(lower) BRE median than Bucket System, while for Sprint 2, Bucket System has a non-statistically significantly 
better (lower) BRE median. Combined data for all four Sprints show that both methods (Planning Poker and Bucket 
System) achieved the same BRE median accuracy of $0.71$.

Thus, we can reject the first H0 since Affinity Estimation is statistically significantly less accurate in 
estimating effort to complete project user stories between studied agile estimation methods. While the 
differences in BRE median accuracy (Table 3) between Planning Poker and Bucket System are too small to state 
that one or the other is a clear winner when comparing them in terms of accuracy.

\begin{table}[h!]
\small
\begin{center}
\begin{tabular}{ @{}llcccll@{} }
\hline\hline 
 & Method accuracy  & Mann-Whitney U & Wilcoxon W & Z & P-value    & Cliff\rq s \\
 & comparison       &                &            &   & (2-tailed) & Delta   \\
\hline
         & BREpp vs BREbs & \hphantom{00}1025.000\hphantom{0} & \hphantom{0}2201.000 & -0.931 & 0.352 & -0.108 \\
Sprint 1 & BREpp vs BREae & \hphantom{00}933.000              & \hphantom{0}2109.000 & -1.605 & 0.108 & -0.191 \\
         & BREbs vs BREae & \hphantom{00}1094.000\hphantom{0} & \hphantom{0}2270.000 & -0.425 & 0.671 & -0.053 \\
\hline
         & BREpp vs BREbs & \hphantom{00}1102.500\hphantom{0} & \hphantom{0}2278.500 & -0.363 & 0.717 & \hphantom{-}0.043 \\
Sprint 2 & BREpp vs BREae & \hphantom{00}941.000              & \hphantom{0}2117.000 & -1.548 & 0.122            & -0.182 \\
         & BREbs vs BREae & \hphantom{00}911.500              & \hphantom{0}2087.500 & -1.764 & 0.078            & -0.209 \\
\hline
         & BREpp vs BREbs & \hphantom{0}1060.500 & \hphantom{0}2236.500 & -0.671 & 0.502 & -0.080 \\
Sprint 3 & BREpp vs BREae & \hphantom{0}1029.000 & \hphantom{0}2205.000 & -0.902 & 0.367 & -0.106 \\
         & BREbs vs BREae & \hphantom{0}1113.000 & \hphantom{0}2289.000 & -0.286 & 0.775 & -0.034 \\
\hline
         & BREpp vs BREbs & \hphantom{0}1079.000 & \hphantom{0}2255.000 & -0.535 & 0.592 & -0.062 \\
Sprint 4 & BREpp vs BREae & \hphantom{0}1005.500 & \hphantom{0}2181.500 & -1.074 & 0.283 & -0.128 \\
         & BREbs vs BREae & \hphantom{0}1093.000 & \hphantom{0}2269.000 & -0.433 & 0.665 & -0.053 \\
\hline
            & BREpp vs BREbs & 17433.500 & 35961.500 & -0.918 & 0.358 & -0.053 \\
All Sprints & BREpp vs BREae & 15601.000 & 34129.000 & -2.604 & 0.009 & -0.154 \\
            & BREbs vs BREae & 16800.500 & 35328.500 & -1.501 & 0.133 & -0.089 \\
\hline\hline
\end{tabular}
\caption{Statistics for statistical testing of accuracy of estimated effort to complete project user stories 
between studied agile estimation methods of PP, BS and AE.}
\label{table_statistics_1}
\end{center}
\end{table}

\subsection{Testing for hypothesis H0: There is no statistically significant difference in time needed to 
estimate project user stories between studied agile estimation methods of PP, BS and AE.} 

To test the second H0, we compared all three studied estimation methods, PP, BS and AE, to each other for 
all Sprints in aggregate only since we had only 8 estimation times per Sprint per method. As can 
be seen in Table 6, there are two statistically significant differences in time (minutes) used between the 
three studied methods. The first is a statistically significant difference between minutes used to conduct 
a sprint estimation with Planning Poker and a sprint estimation with Bucket System with a large Cliff's Delta 
effect size of $0.537$. From Table 4, we can see that it takes a median of $24.5$ minutes for a student team 
to finish a sprint estimation with Planning Poker and 15 minutes to finish a sprint estimation with Bucket 
System; we can thus reject the second H0 and state that Planning Poker is statistically significantly less 
time efficient than Bucket System. The second statistically significant difference is between Planning Poker 
(median time used $24.5$ minutes) and Affinity Estimation (median time used 16 minutes) with again a large 
Cliff's Delta effect size of $0.566$. Thus, we can similarly reject the second H0 and state that Planning 
Poker is also statistically significantly less time-efficient than Affinity Estimation. The comparison between 
Bucket System and Affinity Estimation does not enable us to statistically significantly reject the second H0, 
which is to be expected since their median time used for a sprint estimation differs by less than 
10\% (Table 4).

\begin{table}[h!]
\small
\begin{center}
\begin{tabular}{ @{}llcccll@{} }
\hline\hline 

 & Method time   & Mann-Whitney U  & Wilcoxon W & Z & Asymp.     & Cliff\rq s \\
 & efficiency    &                 &            &   & Sig.       & Delta      \\
 & comparison    &                 &            &   & (2-tailed) &            \\
\hline
            & time used for PP vs  & 237.000 & 765.000 & -3.697 & 0.000  & 0.537 \\
            & time used for BS     &         &         &        &        &       \\						
All         & time used for PP vs  & 222.000 & 750.000 & -3.898 & 0.000  & 0.566 \\
Sprints     & time used for AE     &         &         &        &        &       \\
            & time used for BS vs  & 481.000 & 1009.000\hphantom{0} & -0.417 & 0.677 & 0.061 \\
            & time used for AE     &         &                      &        &       &       \\			

\hline\hline
\end{tabular}
\caption{Statistics for statistical testing of time needed to estimate project user stories in each Sprint between studied 
agile estimation methods of PP, BS and AE.}
\label{table_statistics_2}
\end{center}
\end{table}

Thus, the second research question can be answered with the statement that Planning Poker is the least 
time-efficient estimation method, with a 63\% larger median time needed to complete one estimation when 
compared to Bucket System (Table 3), while the differences in time needed between Affinity Estimation and 
Bucket System are too small to state that one or the other is a clear winner when comparing them in terms 
of time efficiency. The overall winner of the accuracy and time efficiency comparison is thus the agile 
estimation method Bucket System since it has similar accuracy to Planning Poker and better accuracy than 
Affinity Estimation while also being considerably more time efficient than Planning Poker and similarly 
time efficient to Affinity Estimation.

\section{Discussion\label{discussion}}

In recent years, Planning Poker has been used as a de facto standard group effort estimation technique for 
Scrum. However, new consensus-based group estimation techniques are emerging, challenging Planning Poker. 
Mol{\o}kken-{\O}stvold et al. \cite{Molokken-08} mentioned the lack of studies comparing Planning Poker with 
more structured techniques in their work. Our study addresses this issue by comparing the accuracy and time 
efficiency of Planning Poker with two agile estimation techniques that have been gaining popularity in 
recent years. To the best of our knowledge, these two alternative estimation methods, Bucket System and 
Affinity Estimation, were not yet compared with Planning Poker in a case study. Thus, such comparison presents 
our theoretical contribution to the advancement of agile estimation methods evaluation. Another important 
theoretical contribution of our paper is the addition of the second dimension of evaluation. We advanced 
the standard comparison of methods through accuracy \cite{Mahnic-12,Molokken-08,Moharreri-16} by adding the 
additional dimension of time efficiency. We advocate that such evaluation should become standard in the 
field since, in practice, time efficiency is also a very important criterion, especially when the compared 
methods don't differ in accuracy.

The presented evaluation of estimation techniques in a capstone course highlights assessment techniques'\
strengths and deficiencies, which may provide practical benefits in the software engineering and education 
community. The study provides non-trivial and relevant insight into agile estimation techniques. It enables 
lecturers to make informed decisions for improving software engineering courses regarding the time efficiency 
and accuracy of evaluated estimation methods.

Concerning RQ1, statistical tests revealed that Affinity Estimation is statistically significantly less 
accurate than Planning Poker. We also found no statistically significant difference between Bucket System 
and Planning Poker method. However, Affinity Estimation is less accurate than both Bucket System and Planning 
Poker and, therefore, less suitable for estimation. Therefore, we suggest using Planning Poker or Bucket System 
instead of Affinity Estimation for the accuracy of effort estimation. When we look at the papers that 
investigated Planning Poker in terms of its accuracy, the results indicate that several papers found Planning 
Poker to be a superior method. Such results are also in line with our findings, where Planning Poker showed 
itself as a very accurate method. Haugen \cite{Haugen-06} explored whether the Planning Poker estimation 
process yields improved performance compared to unstructured group estimation. They found that Planning Poker 
is more accurate when the team has previous experience with similar tasks. However, teams without previous 
experience might achieve less accurate estimates. Mol{\o}kken-{\O}stvold et al. \cite{Molokken-08} concentrated 
on a comparison between Planning Poker-based group consensus estimates and the statistical combination of 
individual expert estimates. They found that Planning Poker estimates were more accurate than the statistical 
combination of individual estimates for the same tasks. In their case study, Gandomani et al. \cite{Gandomani-14} 
investigated the effectiveness of the Planning Poker and Wideband Delphi. They found that Planning Poker 
estimates were more accurate than Wideband Delphi estimates. Because Planning Poker practice can be 
time-consuming, Gandomani et al. \cite{Gandomani-19} investigated whether the consensus part of Planning Poker 
might be omitted without hindering the estimation accuracy. They compared Planning Poker with a practice 
where an average of suggested points was used and found that Planning Poker yielded higher accuracy estimates. 
On the other hand, Po\v{z}enel and Hovelja \cite{Pozenel-19} compared Planning Poker with the Team Estimation 
Game and found that the Team Estimation Game produces more accurate story estimates than Planning Poker. It 
seems reasonable to continue the research on group estimation techniques in different settings to find 
contexts in which they outperform others.

Regarding RQ2, we were interested in the methods' time efficiency. Planning Poker appeared to be the most 
time-consuming method among those under investigation. Statistical tests found no statistically significant 
differences in time efficiency between Bucket System and Affinity Estimation method, meaning that Bucket System 
is similarly time efficient as Affinity Estimation. However, Bucket System is statistically significantly more 
time efficient than Planning Poker. In other words, Planning Poker is proven to be a more time-consuming method 
than both Bucket System and Affinity Estimation. Assuming that Planning Poker is a consensus-based technique 
based on intensive communication and is known for its time-consuming potential \cite{Gandomani-19,Alhamed-21,Power-11,Pozenel-16}, 
the result is somewhat expected. From this point of view, our results are in line with the papers in the 
literature that argue that Planning Poker can be time-consuming to perform \cite{Pozenel-19,Pozenel-16}. 
Gandomani et al. \cite{Gandomani-19} state that reaching group consensus on the user story size in Planning 
Poker can be difficult and time-consuming in practice. In their work, Alhamed and Storer \cite{Alhamed-21} 
investigated the use of Planning Poker in crowdsourcing. They found that Planning Poker can be time-consuming 
and impractical for large projects requiring assessing thousands of outstanding tasks. Power \cite{Power-11} 
studied the silent grouping technique and discussed that using Planning Poker teams can get stuck in a protracted, 
unnecessary discussion. To fight the issue of the time-consuming estimation process, several authors offered 
algorithmic estimation methods for agile software development \cite{Moharreri-16,Owais-16}.

This study shows that Bucket System can be a viable alternative to Planning Poker since it has similar accuracy 
to Planning Poker and better accuracy than Affinity Estimation while also being more time efficient than Planning 
Poker and similarly time efficient to Affinity Estimation. As an assessment method, Bucket System is just 
gaining ground; thus, there are not many papers in the literature investigating it. The Web site Tech Agilist 
\cite{techagilist-bs} states that it is a good alternative to Planning Poker since it is time-efficient and 
provides reasonably accurate estimates. Our results are in line with findings reported by Tech Agilist 
\cite{techagilist-bs}.

Regarding the changes in the performance of the methods in terms of accuracy and efficiency across Sprints, 
there is no statistically significant improvement or worsening in student precision. In our opinion, this is 
expected since the study was conducted in a master's program where students' overall body of experience in 
evaluating user stories would not be significantly affected by just the experience in the course.

\section{Limitations \label{limitations}}

Empirical studies with students (ESWS) are by some viewed as studies that lack external validity, so the 
value of ESWS for professionals can be under question. Carver et al. \cite{Carver-10} identified the requirements 
for such studies that have to be met in order for them to bring value to students, professionals, and the 
research community. They presented a checklist that helps researchers with all the necessary steps when 
conducting an ESWS to improve the validity of the study. We designed and conducted our study with such 
requirements in mind and followed the checklist presented by Carver et al. \cite{Carver-10}.

Regarding internal validity, we monitored all factors of internal validity to eliminate bias and the 
effects of non-essential variables. All student teams were working on the same project at the same time. 
While conducting the experiment, no changes were introduced in the project settings. Students' participation 
in the study was voluntary and integrated into the course schedule to improve motivation and minimize the 
mortality threat. Students who were not interested in participation had the same study assignments and 
the same curriculum. Changes in the composition of a development team were a part of the simulation of 
changes in an actual project.

Before the study execution, all student teams had the same training concerning software tools and the data 
collection process. They previously also attended the same key classes necessary for the project execution. 
All teams used the same tools for data collection. To mitigate the instrumentation threat, data were collected 
timely and automatically. Students did not know the goals of the study in order not to inadvertently help 
validate the research hypothesis. Also, the outcome of the study did not affect the student's grade. We used 
the Mann-Whitney-Wilcoxon test to compare the observable variables of the student teams. Based on the study 
design and the results of the study, we believe that the only important variable was the assessment method used.

Focusing on external validity, we minimized the threats resulting from the use of ESWS \cite{Host-00} in a 
systematic way to enable a valid generalization. We set the study setting similar to the real software 
environment so that the results could be valuable to the industrial and research communities. To maximize 
the external validity of our study, we faithfully followed the requirements presented by Carver et al. 
\cite{Carver-10} during the study design and execution phases.

Even though Labedzki et al. \cite{Labedzki-17}, when examining the performance of Planning Poker in multiple 
projects with diverse characteristics, found that "Planning Poker is a technique which yields good results in 
any undertaking, increasing the estimation precision and raising the level of understanding user stories which 
await their realization", as far as generalization is concerned, our results are limited to projects with a 
relatively small number of user stories and smaller development teams. The question is whether these three 
methods would yield similar results with larger development teams with significantly larger set of user stories 
in the project. The results can be generalized for students in the last year of undergraduate studies in 
computer science or students in the first year of postgraduate computer science studies. Students in the first 
semester of postgraduate university studies possess theoretical and practical knowledge from the computer 
science field, and many of them have already experience working in software companies on software development 
projects. Thus, they had all the knowledge necessary to participate in the study. We estimate that for 
professionals in software companies, trends in the results would be similar for assessment methods under study. 
We believe we can generalize our findings for tasks related to the team effort estimation needed to complete 
the project.

Concerning construct validity, we ensured that theoretical constructs were interpreted and measured correctly 
\cite{Shull-07}. To minimize the potential danger of students guessing and confirming the hypothesis, we based 
student grades on how faithful students' work complied with the capstone course requirements. We informed 
students that their grade depends on the conformance to the development process and data quality, not the 
study outcome. To avoid an unintentional influence on the students' behaviour, the researchers did not interact 
with the students about the study goals. In agile software development, the assessment process is considered a 
waste \cite{Prause-12}, so we simulated pressure and limited available assessment time, mimicking the assessment 
process of development teams in software companies.

Student teams worked on the same project, resulting in directly comparable estimates, which also increased 
construct validity. We used proven and tested tools like ACScrum, Kanbanize \cite{Mahnic-18}, SPSS and Microsoft 
Excel to collect and process data. The researchers understand the university environment where the study was 
conducted and also benefit from experience with the studies that were conducted in the past \cite{Pozenel-19,Mahnic-15}. 
We believe that we adequately addressed threats to construct validity.

To derive statistically correct conclusions from the collected data, we used a balanced measure of relative 
error (BRE) to calculate estimation accuracy and method time efficiency. To analyze the differences between 
the distributions (BE, AE, and PP), we used non-parametric Mann-Whitney-U tests. Further, we addressed nine 
threats to conclusion validity, presented by Mathay et al. \cite{Matthay-20}. For each threat, we describe 
how we mitigated the threat.

\begin{itemize}
    \item Low statistical power: to mitigate this threat, we additionally report the non-parametric effect 
		size power of Cliff's Delta.
    \item Violated assumptions of statistical tests: we mitigate this by using non-parametric testing with 
		sample sizes larger than $30$.
    \item Fishing and the error rate problem: in the analysis, we included all available data and reported 
		the effect size power of the findings.
    \item Inaccurate effect size estimation: we used published best practices in selecting the non-parametric 
		effect size measure suggested by the literature \cite{Macbeth-11cliff,Goedhart-16calculation}.
    \item Extraneous variance in the experimental setting: The experimental setting followed an established 
		protocol and is repeated yearly.
    \item Heterogeneity of units: all the students are master study students of Computer and Information 
		Science at the University of Ljubljana.
    \item Unreliability of measures: reporting protocols and teams double checking after each other mitigated 
		risks of unreliability.
    \item Restriction of range: the ranges of the variables were not restricted and were in accordance with 
		the requirements of the studied methods.
    \item Unreliability of treatment implementation: methods were implemented in accordance with the best 
		practices described in the literature.
\end{itemize}

We created, tested, and documented the study design before we conducted the study, as recommended by Carver 
et al. \cite{Carver-04}. The resulting study design could be used as a means for study replication in an 
industrial environment or as a stepping stone for further research. We believe we addressed threats to 
statistical conclusion validity that are important to this study.

We also increased the repeatability and reliability of the study with a well-defined case study protocol 
based on a checklist for conducting ESWS \cite{Carver-10,Pozenel-16}, automatic data collection, careful 
preparation and monitoring of study steps.

\section{Conclusions \label{conclusion}}

This paper presents the results of an empirical study on three effort estimation techniques: Bucket System, 
Affinity Estimation and Planning Poker. Although the Planning Poker technique is widely used in Scrum projects 
for effort estimation, new estimation techniques are emerging that attempt to mitigate the deficiencies of 
Planning Poker. Thus, we compared two assessment techniques that have not yet been widely investigated in the 
literature with Planning Poker. The main goal of the paper was to investigate the dimensions of accuracy and 
time efficiency of the studied assessment techniques.

The results of our study show that Affinity Estimation is the least accurate estimation method among the three 
compared. Concerning time efficiency, the results show that Planning Poker is the most time-consuming method 
of the three investigated. The results also show that Bucket System appears to be an attractive assessment 
technique because it has similar accuracy to Planning Poker and better accuracy than Affinity Estimation (RQ1) 
while being more time efficient than Planning Poker and similarly efficient as Affinity Estimation (RQ2).

Although our findings provide new information about studied agile group assessment techniques, there were 
certain limitations to the study. The study had a relatively small sample size with respect to statistical 
analysis and was conducted in a capstone course, so it must be interpreted with care. Considering the aforementioned 
limitations, the results can be generalized to the population of students and young professionals in Computer 
Science and Informatics. The results can also be valuable for lecturers of software engineering courses to 
help them select which evaluation methods should be incorporated into the curriculum. The results may help 
enterprises pick the more time-efficient agile estimation method when the estimation accuracy between methods 
does not differ. It may also be valid for professionals who find Planning Poker too time-consuming but still 
need accurate assessments. Ideally, these findings should be replicated in a study with experienced professionals, 
which would give additional validity to our findings.

We hope our paper will motivate other researchers to replicate the study in different settings to confirm or 
complement our findings. It would be important to replicate the study with industry practitioners, employing 
a significantly larger set of user stories and larger development teams within several projects of diverse 
characteristics. Future research could also focus on the adoption of various automated agile estimation 
approaches compared to group estimation techniques. Further, we believe that it is worth investigating how 
the studied methods compare to the Team Estimation Game assessment technique.

\newpage


\noindent \textbf{For the purpose of open access, the author(s) has applied a Creative Commons attribution (CC BY) 
licence (where permitted by UKRI, ‘Open Government Licence’ or ‘Creative Commons attribution no-derivatives (CC BY-ND) 
licence’ may be stated instead) to any Author Accepted Manuscript version arising.}

\newpage

\bibliographystyle{plain}
\bibliography{paper}

\appendix

\end{document}